\documentclass[aps,prd,twocolumn,floatfix,nofootinbib,showpacs,superscriptaddress]{revtex4-1}
\usepackage{graphicx,amsmath,amssymb,bm}
\usepackage{color}

\definecolor{purple}{rgb}{0.5,0,0.5}
\definecolor{blue}{rgb}{0.0,0,0.9}

\newcommand{\tr}{\mathrm{tr}_\mathrm{CD}\!\ }

\begin{document}

\title{Flavor \textit{SU}(4) breaking between effective couplings}

\author{Bruno~El-Bennich}
\affiliation{Universidade Cruzeiro do Sul, Rua Galv\~ao Bueno, 868, 01506-000 S\~ao Paulo, SP, Brazil}
\affiliation{Instituto de F\'isica Te\'orica, Universidade Estadual Paulista, Rua Dr.~Bento Teobaldo Ferraz, 271, 01140-070 S\~ao Paulo, SP, Brazil}

\author{Gast\~ao Krein}
\affiliation{Instituto de F\'isica Te\'orica, Universidade Estadual Paulista, Rua Dr.~Bento Teobaldo Ferraz, 271, 01140-070 S\~ao Paulo, SP, Brazil}

\author{Lei Chang\,}
\affiliation{Physics Division, Argonne National Laboratory, Argonne, Illinois 60439, USA}

\author{Craig D.~Roberts}
\affiliation{Physics Division, Argonne National Laboratory, Argonne, Illinois 60439, USA}
\affiliation{Institut f\"ur Kernphysik, Forschungszentrum J\"ulich, D-52425 J\"ulich, Germany}
\affiliation{Department of Physics, Illinois Institute of Technology, Chicago, Illinois 60616-3793, USA}

\author{David J.\, Wilson\,}
\affiliation{Physics Division, Argonne National Laboratory, Argonne, Illinois 60439, USA}

\date{4 November 2011}

\begin{abstract}
Using a framework in which all elements are constrained by Dyson-Schwinger equation studies in QCD, and therefore incorporates a consistent, direct and simultaneous description of light- and heavy-quarks and the states they constitute, we analyze the accuracy of $SU(4)$-flavor symmetry relations between $\pi \rho \pi$, $K \rho K$ and $D\rho D$ couplings.  Such relations are widely used in phenomenological analyses of the interactions between matter and charmed mesons.  We find that whilst $SU(3)$-flavor symmetry is accurate to 20\%, $SU(4)$ relations underestimate the $D\rho D$ coupling by a factor of five.  

\end{abstract}

\maketitle

%
\hspace*{-\parindent}\textbf{I.~Introduction}.\hspace*{\parindent}Hadrons in-medium are the focus of intense theoretical and experimental activity.  The chief motivation in heavy-ion collisions is a better understanding of QCD's deconfined phase, {\em viz.\/} the putative quark-gluon plasma, its chiral restoration phase transition and associated order parameters.  Whilst an enhancement of charm and strangeness in the quark-gluon phase is predicted to lead to the copious production of $D_{(s)}$ mesons \cite{Cacciari:2005rkKuznetsova:2006bh} at the large hadron collider, $J/\psi$ suppression has long been suggested as an unambiguous signature for quark-gluon plasma formation \cite{Matsui:1986dk}. Notwithstanding ongoing debates about charmonia production mechanisms and a wide range of suppression effects, much effort is sensibly dedicated to understanding the complicated final-state interactions which occur after hadronization of the plasma; see, e.g., Ref.\,\cite{Bracco:2011pg}.

Charmed-meson interactions with nuclear matter will also be studied at the future Facility for Antiproton and Ion Research (FAIR) and possibly at Jefferson Laboratory (JLab).  Low-momentum charmonia, such as $J/\psi$ and $\psi$, and $D^{(*)}$ mesons can be produced by annihilation of antiprotons on nuclei (FAIR) or by scattering electrons from nuclei (JLab).  Since charmonia do not share valence quarks in common with the surrounding nuclear medium, proposed interaction mechanisms include: QCD van der Waals forces, arising from the exchange of two or more gluons between color-singlet states \cite{Peskin:1979vaBrodsky:1989jd}; and intermediate charmed hadron states \cite{Brodsky:1997ghKo:2000jx}, such that $\bar D^{(\ast)} D^{(\ast)}$ hadronic vacuum polarization components of the $J/\psi$ interact with the medium via meson exchanges \cite{Krein:2010vp}.

A kindred approach is applied to low-energy interactions of open-charm mesons with nuclei, which may create a path to the production of charmed nuclear bound states ($D$-mesic nuclei) \cite{Tsushima:1998ru,Haidenbauer:2007jq,Haidenbauer:2010ch,Yamaguchi:2011xb}.  These studies rely on model Lagrangians, within which effective interactions are expressed through couplings between $D^{(\ast)}$- and light-pseudoscalar- and vector-mesons.  The models are typically an $SU(4)$ extension of light-flavor chirally-symmetric Lagrangians.  Most recently, exotic states formed by heavy mesons and a nucleon were investigated, based upon heavy-meson chiral perturbation theory \cite{Yamaguchi:2011xb}.  In that study a universal coupling, $g_\pi$, between a heavy quark and a light pseudoscalar or vector meson was inferred from the strong decay $D^*\to D\pi$, {\em cf.\/} Ref.\,\cite{ElBennich:2010ha}.

In the context of chiral Lagrangians, it is natural to question the reliability of couplings based on $SU(4)$ symmetry.  Flavor breaking effects are already known to occur in the strange sector and should only be expected to increase when including charm quarks.  The order of magnitude of this larger symmetry breaking is signalled by the compilation of charmed couplings in Ref.\,\cite{Bracco:2011pg}, where $SU(4)$ relations are shown to be violated at various degrees (ranging from 7\% to 70\%) in couplings between two heavy mesons and one light meson.  No states containing a $s$-quark were considered.

Herein, we study a different quantitative measure, based upon ratios between the $D\rho D$, $K\rho K$ and $\pi \rho \pi$ couplings; namely, a difference between the same coupling involving either a $c$-, $s$- or light-quark.  We are motivated by the notion that the $K\rho K$ and $D\rho D$ systems are dynamically equivalent in the sense that the heavier quark acts as a spectator and contributes predominantly to the static properties of the mesons, whereas the exchange dynamics is mediated by the light quarks.  In practice, the symmetry idea is expressed by implementing $g_{D\rho D} \simeq g_{K\rho K}$ in the meson-exchange models \cite{Haidenbauer:2007jq,Haidenbauer:2010ch}.  The $\pi \rho \pi$ coupling provides a well-constrained benchmark.

\hspace*{-\parindent}\textbf{II.~DSE Framework}.\hspace*{\parindent}Our primary object of interest is a phenomenological coupling that relates the transition amplitude of an initial pseudoscalar $H=Qf$-meson, $Q=c,s$ and $f=u,d$, to an identical meson via emission of an off-shell $\rho$.  The matrix element for this transition is
\begin{equation}
  \langle  H(p_2) | \, \rho (P,\lambda)\,  | H (p_1 ) \rangle = g_{H \rho H} \  \bm{\epsilon}_\lambda\!\cdot P \, ,
\label{eq1}
\end{equation}
an expression which defines the dimensionless coupling of the two pseudoscalar mesons to a vector meson with momentum $P=p_2-p_1$ and polarization state $\lambda$.  The decay $\rho\to \pi\pi$ is also described by such a matrix element.  However, there is no associated physical process when $m_\rho^2<4 m_H^2$ and $p_1^2 = p_2^2=-m_H^2$.  (N.B.\ A Euclidean metric is used: $\{\gamma_\mu, \gamma_\nu \} =2\,\delta_{\gamma\nu};\,  \gamma_\mu^\dagger = \gamma_\mu; \; a\!\cdot\! b = \sum^4_{i=1} a_i b_i$; and $\mathrm{tr} [\gamma_5\gamma_\mu\gamma_\nu\gamma_\rho \gamma\sigma ] = -4 \epsilon_{\mu\nu\rho\sigma}, \, \epsilon_{1234} = 1$.  For a space-like vector $P_\mu, P^2 > 0$.)
Nevertheless, a coupling of this sort is employed in defining $\rho$-meson-mediated exchange-interactions between a nucleon and pseudoscalar strange- or charm-mesons.  In such applications: the off-shell $\rho$-meson's momentum is necessarily spacelike; and a coupling and form factor may be defined once one settles on a definition of the off-shell $\rho$-meson.

Symmetry-preserving models built upon predictions of QCD's Dyson-Schwinger equations (DSEs) provide a sound framework within which to examine heavy-meson observables \cite{ElBennich:2010ha,Ivanov:1997yg,Ivanov:1998ms,Ivanov:2007cw,ElBennich:2009vx}.  Such studies describe quark propagation via fully dressed Schwinger functions, which has a material impact on light-quark characteristics \cite{Chang:2011vu}.

At leading-order in a systematic, symmetry-preserving truncation scheme \cite{Bender:1996bb}, one may express Eq.\,(\ref{eq1}) as
\begin{eqnarray}
  g_{H\!\rho H} \  \bm{\epsilon}^\lambda\!\cdot P  & =  &  \tr \! \int\! \frac{d^4k}{(2\pi)^4} \, \Gamma_H (k;k_1)  S_Q(k_Q) \nonumber   \\
  &  &\hspace*{-1.5cm} \times \  \bar  \Gamma_H(k;-k_2) S_f(k_f')\, \bm{\epsilon}^{\lambda*}\!\cdot \bar \Gamma_\rho(k;-P) S_f(k_f) \;  ,
\label{eq2}
\end{eqnarray}
where $S$ represent dressed-quark propagators for the indicated flavor and $\Gamma_H$ are meson Bethe-Salpeter amplitudes (BSAs), with $H=\pi, K, D$.  In Eq.\,(\ref{eq2}): the trace is over color and spinor indices;  $k_Q = k+w_1 p_1,  k_f' =k+w_1 p_1 - p_2$, $k_f =k-w_2p_1$, where the relative- momentum partitioning parameters satisfy $w_1 + w_2 = 1$; and $\bm{\epsilon}^\lambda_\mu$ is the vector-meson polarization four-vector.  This approximation has been employed successfully; see, for instance, applications in Refs.\,\cite{Ivanov:2007cw,Chang:2011vu,Roberts:1994hh,Tandy:1997qf,Jarecke:2002xd,%
Maris:2003vk,Roberts:2007jh}.

We simultaneously calculate the $D$-, $K$- and $\rho$-meson leptonic decay constants via  \cite{Ivanov:1998ms}:
\begin{eqnarray}
\label{psdecay}
P_\mu f_{H}  &=& \tr \int\!  \frac{d^4k}{(2\pi)^4} \, \gamma_5 \gamma_\mu\, \chi_{H}(k;P)\,, \\
\label{vecdecay}
M_\rho f_\rho & = & \frac{1}{3}\tr \int\!  \frac{d^4k}{(2\pi)^4} \,  \gamma_\mu\, \chi_\mu^\rho (k;P) \, ,
\end{eqnarray}
where $\chi(k;P) = S_{f_1}(k+w_1P) \Gamma(k;P) S_{f_2}(k-w_2P)$.  The BSAs are canonically normalized; {\em viz\/}., for pseudoscalars
\begin{eqnarray}
  2\, P_\mu & = & \left [ \frac{\partial}{\partial K_\mu} \Pi(P,K) \right ]_{K=P}^{P^2=-m^2_{0^-}} \ ,
  \label{norm1} \\
   \Pi(P,K) & = &   \tr \int\!  \frac{d^4k}{(2\pi)^4} \, \bar \Gamma_{0^-}(k;-P) S_{f_1}(k+w_1K) \nonumber \\
   &  & \times \   \Gamma_{0^-}(k;P) S_{f_2}(k-w_2K) \,,
  \label{norm2}
\end{eqnarray}
with an analogous expression for the $\rho$ \cite{Ivanov:1998ms}.

The solution of QCD's gap equation is the dressed-quark propagator, which has the general form
\begin{equation}
\label{SpAB}
S(p) = -i \gamma\cdot p\, \sigma_V(p^2) + \sigma_S(p^2)= 1/[i\gamma\cdot p\, A(p^2) + B(p^2)] \, .
\end{equation}
For light-quarks, it is a longstanding DSE prediction that both the wave-function renormalization, $Z(p^2)=1/A(p^2)$, and dressed-quark mass-function, $M(p^2)=B(p^2)/A(p^2)=\sigma_S(p^2)/\sigma_V(p^2)$, receive strong momentum-dependent modifications at infrared momenta: $Z(p^2)$ is suppressed and $M(p^2)$ enhanced.  These features are characteristic of dynamical chiral symmetry breaking (DCSB) and, plausibly, of confinement.  (N.B.\ Eqs.\,(\protect\ref{ssm}), (\protect\ref{svm}) represent the quark propagator $S(p)$ as an entire function, which entails the absence of a Lehmann representation and is a sufficient condition for confinement \protect\cite{Krein:1990sf,Roberts:2007ji}.)  The significance of this infrared dressing has long been emphasized \cite{Roberts:1994hh}; e.g., it is intimately connected with the appearance of Goldstone modes \cite{Chang:2011vu}.  The predicted behavior of $Z(p^2)$, $M(p^2)$ has been confirmed in numerical simulations of lattice-regularized QCD \cite{Roberts:2007ji,Bowman:2005vxBhagwat:2006tu}.

Whilst numerical solutions of the quark DSE are readily obtained, the utility of an algebraic form for $S(p)$, when calculations require the evaluation of numerous integrals, is self-evident.  An efficacious parametrization, exhibiting the aforementioned features and used extensively \cite{Ivanov:1998ms,Ivanov:2007cw,Roberts:1994hh,Cloet:2008re}, is expressed via
\begin{eqnarray}
\nonumber \bar\sigma_S(x) & =&  2\,\bar m \,{\cal F}(2 (x+\bar m^2))\\
&&  + {\cal
F}(b_1 x) \,{\cal F}(b_3 x) \,
\left[b_0 + b_2 {\cal F}(\epsilon x)\right]\,,\label{ssm} \\
\label{svm} \bar\sigma_V(x) & = & \frac{1}{x+\bar m^2}\, \left[ 1 - {\cal F}(2 (x+\bar m^2))\right]\,,
\end{eqnarray}
with $x=p^2/\lambda^2$, $\bar m$ = $m/\lambda$, ${\cal F}(x)= [1-\exp(-x)]/x$,
$\bar\sigma_S(x) = \lambda\,\sigma_S(p^2)$ and $\bar\sigma_V(x) =
\lambda^2\,\sigma_V(p^2)$.  The parameter values were fixed \cite{Ivanov:1998ms} by requiring a least-squares fit to a wide range of light- and heavy-meson observables, and take the values:
\begin{equation}
\label{tableA}
\begin{array}{llcccc}
f &   \bar m_f& b_0^f & b_1^f & b_2^f & b_3^f \\\hline
u=d &   0.00948 & 0.131 & 2.94 & 0.733 & 0.185 \\
  s    &   0.210   &  0.105  & 3.18 &  0.858 & 0.185
\end{array} \, .
\end{equation}
At a scale $\lambda=0.566\,$GeV, the current-quark masses take the values $m_u=5.4\,$MeV and $m_s=119\,$MeV, and one obtains the following Euclidean constituent-quark masses \cite{Maris:1997tm}:
$\hat M_u^E = 0.36\,$GeV and $\hat M_s^E = 0.49\,$GeV.  (N.B.\ $\epsilon=10^{-4}$ in Eq.\,(\protect\ref{ssm}) acts only to decouple the large- and intermediate-$p^2$ domains \protect\cite{Roberts:1994hh}.)

We note that studies which do not or cannot implement light-quark dressing in this QCD-consistent manner invariably encounter problems arising from the need to employ large constituent-quark masses and the associated poles in the light-quark propagators \cite{ElBennich:2008xyElBennich:2008qa}.  This typically translates into considerable model sensitivity for computed observables \cite{ElBennich:2009vx}.

Whereas the impact of DCSB on light-quark propagators is significant, the effect diminishes with increasing current-quark mass (see, e.g., Fig.~1 in Ref.\,\cite{Ivanov:1998ms}).  This can be explicated by considering the dimensionless and renormalization-group-invariant ratio $\varsigma_f:=\sigma_f/M^E_f$, where $\sigma_f$ is a constituent-quark $\sigma$-term: $\varsigma_f$ measures the effect of explicit chiral symmetry breaking on the dressed-quark mass-function compared with the sum of the effects of explicit and dynamical chiral symmetry breaking.  Calculation reveals \cite{Roberts:2007jh}: $\varsigma_u = 0.02$, $\varsigma_s = 0.23$, $\varsigma_c = 0.65$, $\varsigma_b = 0.8$.  Plainly, $\varsigma_f$ vanishes in the chiral limit and remains small for light quarks, since the magnitude of their constituent mass owes primarily to DCSB.  On the other hand, for heavy quarks, $\varsigma_f\to 1$ because explicit chiral symmetry breaking is the dominant source of their mass.  Notwithstanding this, confinement remains important for the heavy-quarks.  These considerations are balanced in the following simple form for the $c$-quark propagator:
\begin{equation}
\label{SQ}
  S_c (k) = \frac{-i \gamma\cdot k + \hat M_c}{\hat M_c^2} {\cal F}(k^2/\hat M_c^2)\,,
\end{equation}
which implements confinement but produces a momentum-independent c-quark mass-function; namely, $\sigma_V^c(k^2)/\sigma_S^c(k^2)=\hat M_c$.   We use $\hat M_c = 1.32\,{\rm GeV}$ \cite{Ivanov:1998ms}.

A meson is described by the amplitude obtained from a homogeneous Bethe-Salpeter equation.  In solving that equation the simultaneous solution of the gap equation is required.  Since we have already chosen to simplify the calculations by parametrizing $S(p)$, we follow Refs.\,\cite{ElBennich:2010ha,Ivanov:1998ms,Ivanov:2007cw,ElBennich:2009vx} and also employ that expedient with $\Gamma_{H(\rho)}$.

In this connection, the quark-level Goldberger-Treiman relations derived in Ref.\,\cite{Maris:1997hd} motivate and support the following parametrization of the $\pi$ and $K$ BSAs:
\begin{equation}
\label{piKamp}
  \Gamma_{\pi,K}(k;P) = i\gamma_5\,\frac{\surd 2}{f_{\pi,K}}\,B_{\pi,K} (k^2)\,,\\
\end{equation}
%
where
$B_{\pi,K}:=\left. B_u\right|_{m_u\to 0}^{b_0^u\to b_0^{\pi,K}}$ and
%
%
are obtained from Eqs.\,(\ref{SpAB}) -- (\ref{svm}) through the replacements
$b_0^u \rightarrow b_0^\pi = 0.204$, $b_0^u \rightarrow b_0^K = 0.319$, which yield computed values $f_\pi = 146\,$MeV, $f_K = 178\,$MeV \cite{Ivanov:1998ms}.  Equation~(\ref{piKamp}) expresses the fact that the dominant invariant function in a pseudoscalar meson's BSA is closely related to the scalar piece of the dressed-quark self energy owing to the axial-vector Ward-Takahashi identity and DCSB.


Regarding the $\rho$ meson, DSE studies 
\cite{Jarecke:2002xd,Pichowsky:1999mu} indicate that, in applications such as ours, one may effectively use
\begin{equation}
\label{GV}
 \Gamma^\mu_\rho (k;P) =  \left ( \gamma^\mu -P^\mu\, \frac{\gamma\cdot P}{P^2} \right )   \frac{\exp (-k^2/ \omega_\rho^2) }{\mathcal{N}_\rho}  \, ,
\end{equation}
namely, a function whose support is greatest in the infrared.  Similarly, for the $D$ meson we choose:
\begin{equation}
\label{GH}
   \Gamma_D (k;P) = i \gamma_5 \, \frac{\exp (-k^2/\omega_D^2) }{\mathcal{N}_D}  \; .
\end{equation}

The normalizations, $\mathcal{N}_\rho$, $\mathcal{N}_D$, are obtained from Eqs.\,\eqref{norm1}, \eqref{norm2} and simultaneous calculation
of the weak decay constant in Eqs.\,\eqref{psdecay}, \eqref{vecdecay}.  In the expression for the coupling, Eq.\,\eqref{eq1}, as well as in Eqs.~\eqref{psdecay}--\eqref{norm1}, we follow the momentum-partitioning prescription of Ref.\,\cite{ElBennich:2010ha}, which leads to $w_1^c = 0.79$;
{\em viz\/}., most but not all the heavy-light-meson's momentum is carried by the $c$-quark.
We note that Poincar\'e covariance is a hallmark of the direct application of DSEs to the calculation of hadron properties.  In such an approach, no physical observable can depend on the choice of momentum partitioning.  However, that feature is compromised if, as herein, one does not retain the complete structure of hadron bound-state amplitudes \cite{Maris:1997tm}.  Any sensitivity to the partitioning is an artifact arising from our simplifications \cite{Ivanov:2007cw,ElBennich:2010ha}.

\hspace*{-\parindent}\textbf{III.~Results}.\hspace*{\parindent}%
%
The $D$-meson's width parameter is determined via analysis of relevant leptonic and strong decays: $\omega_D = 1.63\pm 0.10\,$GeV for $m_D=1.865\,$GeV yields $f_D=206\pm 9\,$MeV \cite{Eisenstein:2008sq} and $g_{D^\ast D \pi}=18.7^{+2.5}_{-1.4}$ \emph{cf}. $17.9\pm1.9$ \cite{Anastassov:2001cw}.
For the $\rho$, we use $\omega_\rho =  0.56 \pm 0.01\,$GeV and $w_2^\rho = 0.38$, both determined \cite{Ivanov:2007cw} via a least-squares fit to an array of light-light- and heavy-light-meson observables with $m_\rho=0.77\,$GeV.  Using Eqs.\,\eqref{psdecay}, \eqref{norm1} and \eqref{norm2}, one therewith obtains $f_\rho =209\,$MeV, \emph{cf}.\ experiment $216\,$MeV, which follows from the $e^+ e^-$ decay width \cite{Nakamura:2010zzi}.

\begin{figure}[t]
\centerline{\includegraphics[clip,width=0.4\textwidth]{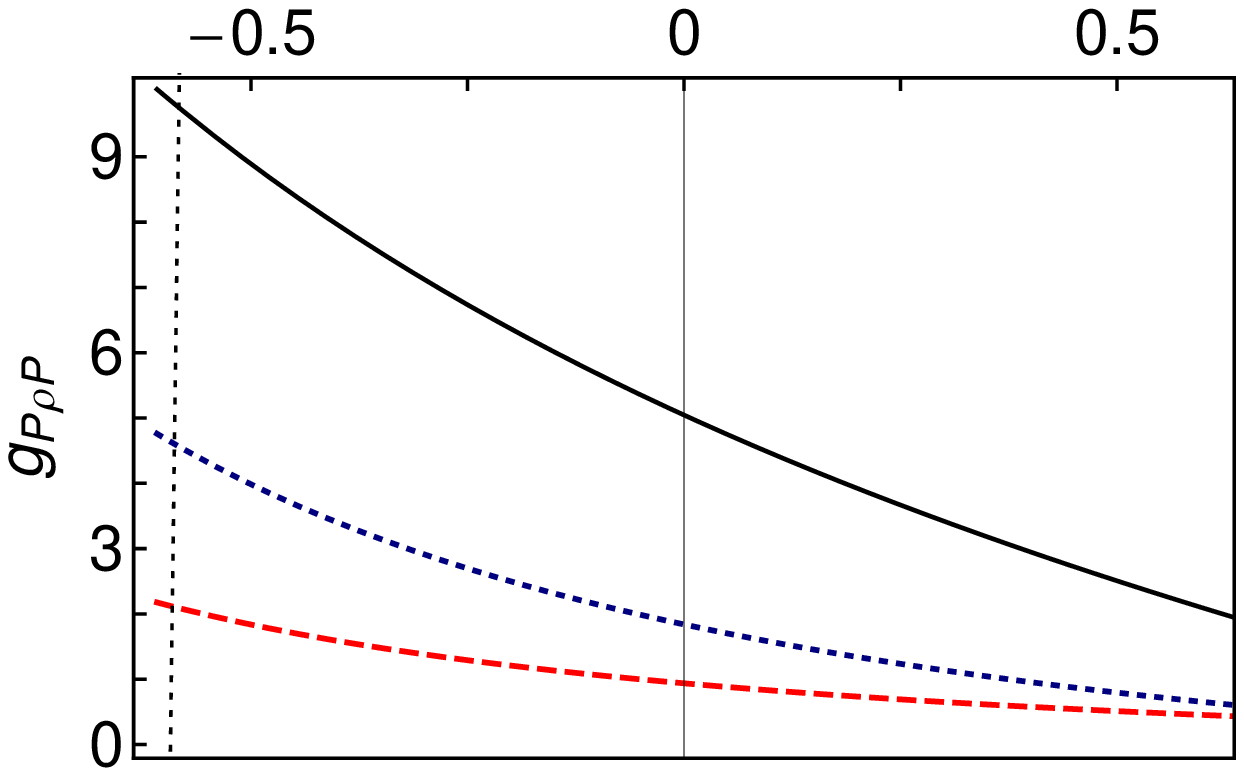}}
\vspace*{-4.4ex}

\centerline{\includegraphics[clip,width=0.403\textwidth]{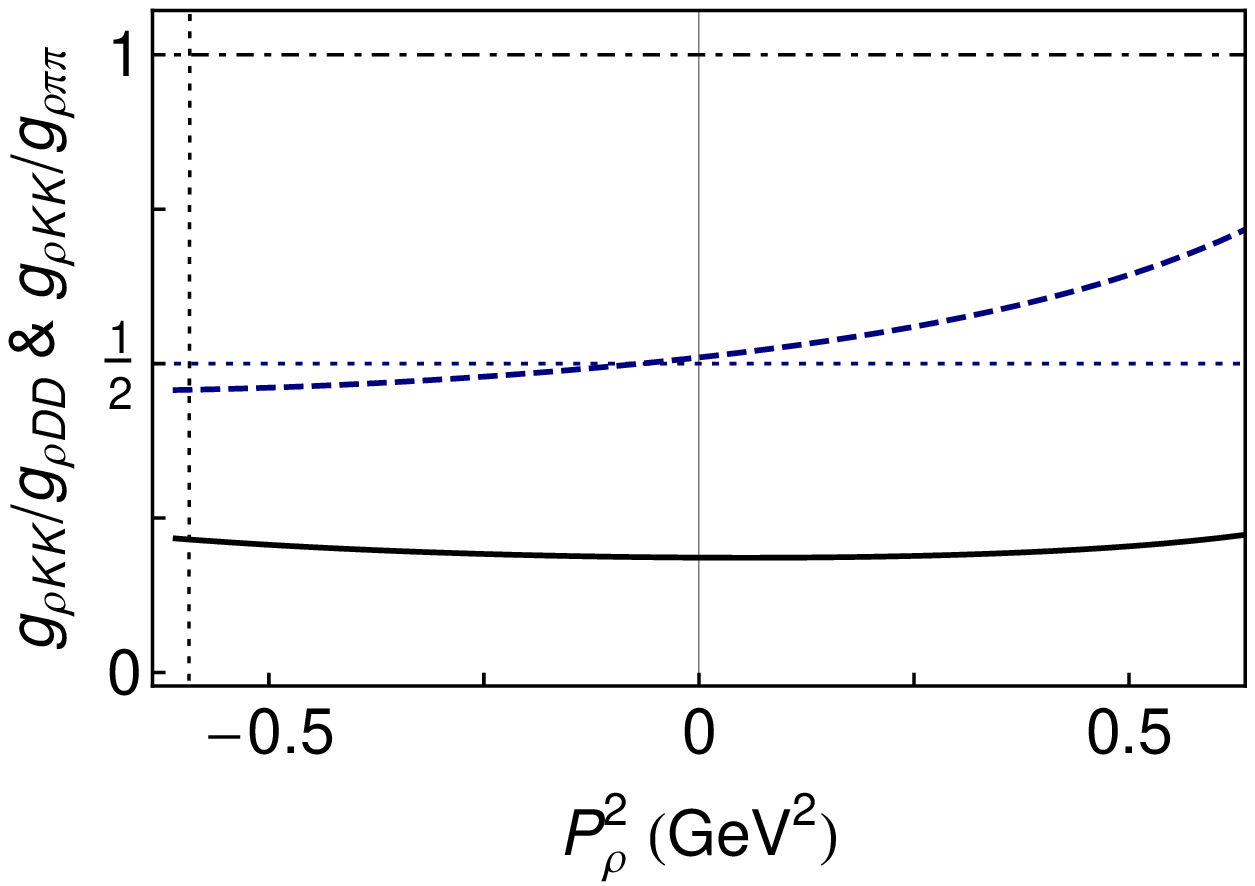}\hspace*{0.15em}}
\caption{
\emph{Upper panel} -- Dimensionless couplings: $g_{D\rho D}$ (solid curve); $g_{K\rho K}$ (dashed curve); and $g_{\pi\rho\pi}$ (dotted curve) -- all computed as a function of the $\rho$-meson's off-shell four-momentum-squared, with the pseudoscalar mesons on-shell.  Recall that with our Euclidean metric, $P^2>0$ is spacelike.
\emph{Lower panel} -- Ratios of couplings: $g_{K\rho K}/g_{D\rho D}$ (solid curve); and $g_{K\rho K}/g_{\pi\rho \pi}$ (dashed curve).  In the case of exact $SU(4)$ symmetry, these ratios take the values, respectively, $1$ (dot-dashed line) and $(1/2)$ (dotted line).
The vertical dotted line marks the $\rho$-meson's on-shell point in both panels.
(N.B.\ In GeV: $m_D =1.865$, $m_\rho=0.77$, $m_K=0.494$, $m_\pi = 0.138$.)
\label{figcoupling}}
\end{figure}

With the width parameters fixed, we computed the $D\rho D$, $K\rho K$ and $\pi \rho \pi$ couplings in impulse approximation, following Eq.\,\eqref{eq2}.  Our results are depicted in Fig.\,\ref{figcoupling}.  Notably, we compute the amplitude directly: at all values of $P^2$ and current-quark mass.  We do not need to resort to extrapolations, neither from spacelike$\,\to\,$timelike momenta nor in current-quark mass, expedients which are necessary in some other approaches \cite{Bracco:2011pg,Becirevic:2009xp}.

The behavior of $g_{\pi\rho\pi}(P^2)$ provides a context for our results.  Experimentally \cite{Nakamura:2010zzi}, $g_{\pi\rho\pi}(-m_\rho^2)=6.0$; and the best numerically-intensive DSE computation available produces \cite{Jarecke:2002xd} $g_{\pi\rho\pi}(-m_\rho^2)=5.2$.  Our algebraically-simplified framework produces $g_{\pi\rho\pi}(-m_\rho^2)=4.8$, just 8\% smaller than the latter, and a $P^2$-dependence for the coupling which closely resembles that in Ref.\,\cite{Mitchell:1996dn}; e.g., both are smooth, monotonically decreasing functions and our value of $g_{\pi\rho\pi}(-m_\rho^2)/g_{\pi\rho\pi}(m_\rho^2)=0.14$ is just 10\% smaller.  On the domain $P^2\in [-m_\rho^2,m_\rho^2]$
\begin{equation}
g_{\pi\rho\pi}(s=P^2) = \frac{1.84 -1.45 s}{1+ 0.75 s + 0.085 s^2}
\end{equation}
provides an accurate interpolation of our result.  If one insists on a monopole parametrization at spacelike-$P^2$, then a monopole mass of $\Lambda_{\pi\rho\pi}=0.61\,$GeV provides a fit with relative-error-standard-deviation$\,=5$\%.

In the case of exact $SU(3)$ symmetry, one would have $g_{K \rho K} = g_{\pi\rho \pi}/2$.  It is clear from the figure that the assumption provides a fair approximation to our result on a domain which one can reasonably consider as relevant to meson-exchange model phenomenology; viz., on $P^2\in [-m_\rho^2,m_\rho^2]$ the error ranges from $(-10)\,$--$40\,$\%.  On this domain an accurate interpolation is provided by
\begin{equation}
\label{IgrKK}
g_{K\rho K}(s) = \frac{0.94 -0.62 s}{1 + 0.55 s - 0.16 s^2}.
\end{equation}
If one insists on a monopole parametrization at spacelike-$P^2$, then a monopole mass of $\Lambda_{K\rho K}=0.77\,$GeV provides a fit with relative-error-standard-deviation$\,=4$\%.

With $SU(4)$ symmetry, the picture is different.  We have a numerical result that is reliably interpolated via
\begin{equation}
g_{D\rho D}(s)=\frac{5.05 -4.26 s}{1+0.36 s- 0.060 s^2}.
\end{equation}
A monopole parametrization at spacelike-$P^2$, with mass-scale $\Lambda_{D\rho D}=0.69\,$GeV, provides a fit with relative-error-standard-deviation$\,=5$\%.  
Our computed value $g_{D\rho D}(0)=5.05$ is 75\% larger than an estimate obtained using QCD sum rules ($3.0\pm 0.02$ \cite{Bracco:2011pg}) and 100\% larger than a vector-meson-dominance estimate ($2.52$ \cite{Lin:1999ad}).  
Moreover, if $SU(4)$ symmetry were exact, then $g_{D\rho D} = g_{K \rho K} = g_{\pi\rho \pi}/2$, but it is plain from Eq.\,(\ref{IgrKK}) that $g_{K\rho K}(0)=0.92$, a result which exposes a symmetry violation of $440$\% at $P^2=0$.  Furthermore, on the entire domain $P^2\in [-m_\rho^2,m_\rho^2]$, the symmetry-based expectation $g_{D\rho D} = g_{K \rho K}$ is always violated, at a level of between $360\,$--$\,440$\%.  The second identity, $g_{D\rho D}=g_{\pi\rho \pi}/2$, is violated at the level of $320\,$--$\,540$\%.  (N.B.\ In connection with heavy-quark symmetry, corrections of this order have also been encountered $c\to d$ transitions \cite{Ivanov:1998ms}.)

These conclusions are dramatic, so it is important to explain why we judge them to be robust.
The computations of $g_{\pi\rho\pi}$ and $g_{K\rho K}$ are considered reliable because we can smoothly take the limit $s$-quark$\,\to\,u$-quark and thereby recover a unique function that agrees with earlier computations by other groups.

This leaves the possibility of uncertainties connected with $S_c(k)$, Eq.\,(\ref{SQ}); $\Gamma_D(k;P)$, Eq.\,(\ref{GH}); and the momentum partitioning parameter, $w_1^c$.
To explore sensitivity to the $c$-quark propagator we used an even simpler, non-confining constituent-like form; viz., $S_C(k)=1/(i\gamma\cdot k + \hat M_c)$.  The effect at spacelike-$P^2$ is modest.  However, the impact is large at timelike-$P^2$ because thereupon the $\rho$-meson momentum-squared begins to explore a neighborhood of the spurious pole in $S_C(k)$.  Thus, the simpler propagator serves to \emph{increase} the violation of $SU(4)$ symmetry.
Regarding $\Gamma_D(k;P)$, uncertainty is implicit in the value of $\omega_D = 1.63\pm 0.10\,$GeV, constrained by the weak decay constant $f_{D^+} =206\pm 9\,$MeV \cite{Eisenstein:2008sq}.  However, variations of even 20\% in $\omega_D$ have no material impact on our results.
Connected with that, a 20\% change in $w_1^c$ produces only a 4\% variation in $\omega_D$ via the fit to $f_{D^+}$, hence any possibility of an effect from $w_1^c$ can be discounted owing to the previous consideration.


\hspace*{-\parindent}\textbf{IV.~Discussion}.\hspace*{\parindent}Predictions for bound-states and resonances derived from meson-exchange models are sensitive to the values of couplings in their Lagrangians.  In these non-relativistic models the couplings are commonly fixed to reproduce some known experimental data, e.g.\ the scattering length of a physical system.  The most prominent such coupling, namely $g_{\pi\!N}$, has long been used in nucleon-nucleon potentials and serves to define the strength of the pion's coupling to a nucleon.  It also determines the scale of the long-range force in the nucleon-nucleon interaction and associated scattering cross sections.  Analogously, the strength of the couplings $D^{(\ast)}\!D\pi$, $D^{(\ast)}D^{(\ast)}\rho$ between $D$ mesons and a light pion or $\rho$-meson plays a crucial role in the formation of charmed-nuclei.  However, whereas $g_{\pi\!N}$ can be extracted from $\pi N$-scattering data \cite{Ericson:2000md}, no such information is available for charmed-meson interactions with nucleons.

In our approach, which is based on an internally consistent use of impulse approximation and unifies the description of light- and heavy-mesons, we compute these couplings
from the transition amplitude between two $D$ mesons and an off-shell light meson.  We find that $SU(4)$ symmetry is a very poor guide to the couplings.  On the other hand, in relation to such models it provides a constructive suggestion that one might reasonably employ
\begin{equation}
\label{FFME}
F^{\rm ME}_{D \rho D}(|\vec{q}|^2) =
g^{\rm ME}_{D \rho D}
\frac{\Lambda_{D \rho D}^{{\rm ME}\,2}}{\Lambda_{D \rho D}^{{\rm ME}\,2}+|\vec{q}|^2},
\end{equation}
with $g^{\rm ME}_{D \rho D} \approx 5$, $\Lambda^{\rm ME}_{D \rho D} \approx 0.7\,$GeV, to describe $D\,D$ scattering via $\rho(\vec{q})$-meson exchange.

This might be compared with the parametrization \cite{Haidenbauer:2007jq}:
\begin{equation}
\label{FFH}
F^H_{D \rho D}(|\vec{q}|^2) = g^H_{D \rho D}\frac{\Lambda^{H\,2}_{D \rho D}}{\Lambda_{D \rho D}^{H\,2}+|\vec{q}|^2},
\end{equation}
$\Lambda^H_{D \rho D}=1.4\,$GeV, $g^H_{D \rho D} \approx 2$, based on the notion of $SU(4)$ symmetry, which our analysis has discredited.
%
The coupling in Eq.\,(\ref{FFH}) is smaller than that in Eq.\,(\ref{FFME}) but the evolution is harder.  These effects cancel to some degree, but here the magnitudes are such that our result, Eq.\,(\ref{FFME}), provides an integrated interaction
\begin{equation}
V_0 = \int d^3 \vec{q} \; F^H_{D \rho D}(|\vec{q}|^2)^2 \frac{1}{|\vec{q}|^2+m_\rho^2}
\end{equation}
that is roughly 40\% greater.  (N.B.\ If $g^H_{D \rho D} \to 2.6 \approx (1/2)g^{\rm ME}_{D\rho D}$, then $V_0^H \approx V_0^{ME}$.)
By the same measure, our $D\rho D$ interaction is 20\% stronger than that in Ref.\,\cite{Yamaguchi:2011xb}, which uses $\Lambda^Y_{D\rho D}=1.14\,$GeV, $g_V=5.8$ and hence
\begin{equation}
g_{D\rho D}^Y= 0.9 g_V [1-m_\rho^2/\Lambda^{Y\,2}_{D\rho D}] = 2.85\,.
\end{equation}%
Whilst our results argue against hard form factors, the interaction enhancement they produce is abundantly clear.  Notably, a large value for the interaction strength entails an inflated cross-section in $D N$ scattering.  In particular, in the meson-exchange model of Ref.\,\cite{Haidenbauer:2007jq} (single-meson exchange version), the $I=1$ $\bar D N$ cross-section is inflated by a factor of $\sim 5$, when using the our result, Eq.\,\eqref{FFME}, for $\omega$ and $\rho$, instead of Eq.\,\eqref{FFH}.  Hence, implementation of our results could have material consequences on, e.g., the possibility for formation of charmed-resonances or -bound-states in nuclei.

\begin{acknowledgments}
We acknowledge useful input from A.~Hosaka and S.\,M.~Schmidt.  This work was supported by: 
Conselho Nacional de Desenvolvimento Cient\'{\i}fico e Tecnol\'ogico, grant no.\ 305894/2009-9, Funda\c{c}\~ao de Amparo \`a  Pesquisa do Estado de S\~ao Paulo, grant nos.\ 2009/50180-0, 2009/51296-1 and 2010/05772-3; United States Department of Energy, Office of Nuclear Physics, contract no.~DE-AC02-06CH11357; and Forschungszentrum J\"ulich GmbH.
\end{acknowledgments}


\end{document}